\documentclass[superscriptaddress,twocolumn,showpacs,aps,prb]{revtex4}
\usepackage{amssymb}
\usepackage{amsmath}
\usepackage{amsfonts}
\usepackage{graphicx}
\usepackage{dcolumn}
\usepackage{natbib,hyperref}
\usepackage{bm}
\usepackage{soul}
\usepackage{epstopdf}
\usepackage{color}

\begin{document}

\title{Numerical analysis of spin-orbit coupled one dimensional Fermi gas in the magnetic field}
\author{Y.H. Chan}
\affiliation{Institute of Atomic and Molecular Sciences, Academia Sinica, Taipei 10617, Taiwan}

\begin{abstract}
We use the density matrix renormalization group method(DMRG) and the infinite time evolved block
decimation method(iTEBD) to investigate the ground states of the spin-orbit coupled Fermi gas in a one dimensional optical lattice with a transverse magnetic field. 
We discover that the system with attractive interaction can have a polarized insulator(PI), a superconducting phase(SC), a Luther-Emery(LE)
phase and a band insulator(BI) phase as we vary the chemical potential and the strength of magnetic field. 
We find that spin-orbit coupling induces a triplet pairing order at zero momentum with the same critical exponent as that of the singlet pairing one in both the SC and the LE phase.
In contrast to the FFLO phase found in the spin imbalanced system without spin-orbit coupling, pairings at finite momentum in these two phases have a larger exponent hence do not dictate the long range behavior.
We also find good agreements of the dominant correlations between numerical results and the prediction from the bosonization method. 
The presence of Majorana fermions is tested.
However, unlike results from the mean field study, we do not find positive evidence of Majorana fermions in our system.

\end{abstract}

\maketitle

\section{Introduction}
Spin-orbit coupling(SOC) in both the condensed matter and the cold atom system has drawn a lot of attention in recent years. It plays a crucial role in the spin Hall effects\cite{Murakami2004}, spintronic devices\cite{Ziuti2004}, and novel topological phases\cite{Jackeli2009,Pesin2010}. In these systems strong SOC modifies the band structure and the band topology. Synthetic SOC in the cold atom system\cite{Dalibard2011,ZHAI2012} also gives rise to novel pairing states such as Flude-Ferrell state, and Larkin-Ovchinnikov state with topological properties, where Majorana fermion is found in the soliton excitation in the theoretical investigation\cite{Wu2013,Zhang2013,Qu2014,Xu2014}. The interplay between SOC and interaction effects has brought interesting physics in both fields.

The development in the cold atom experiments in the optical lattice has provided new insight in one dimensional quantum many-body systems.
In particular there are renewed interest in 1D fermi gas\cite{Guan2013}. Novel pairing states\cite{Sun2014,Zhang2013,Xu2014}, large spin physics\cite{Ho2000}, dimensional crossover\cite{Sun2013} and the dynamical properties has been widely discussed theoretically and experimentally. Thanks to the powerful technique such as Bethe ansatz, bosonization method, and various numerical tools developed for 1D systems a better understanding and a thorough comparison with experiments are possible. 

Recently, great efforts has been focused on spin-orbit coupled 1D fermi gas due to the proposed realization\cite{Oreg2010} of Majorana fermion in the nano wire system. Majorana fermion, which has the application as a basic unit in the fault-tolerant topological quantum computing\cite{Sau2010,Alicea2011}, is predicted in a 1D $p$-wave superconductor\cite{Alexei2000} yet its direct implementation is not achieved so far. The proposed equivalent model consists of the spin-orbit coupled nano-wire in a transverse magnetic field with the pairing potential induced by the proximity effect of an adjacent $s$-wave superconductor.\cite{Oreg2010} Its experimental implementation was soon established and the evidence of Majorana fermions has also been claimed.\cite{Mourik2012} On the cold atom side implementations of SOC was realized in the bosonic systems.\cite{Lin2009,Lin2011}The progress in the fermionic systems is also encouraging. Two groups have successfully implemented the SOC with cold fermions and the single particle spectrum can be clearly observed in their experiments.\cite{Cheuk2012,Wang2012} However, it is not clear if the proximity effect, as a crucial ingredient of Majorana fermions in this proposal can be induced in the cold atom systems. 

In contrast, the attractive interaction tuned by Feshbach resonance seems to be a more straightforward option for the pairing mechanism in the cold atom system. 
Several theoretical studies have shown that in a similar two dimensional model the attractive interaction can stabilize the $p$-$ip$ superconducting 
phases on the mean field level.\cite{Zhang2008,Sato2009} 
For a one dimensional system, it has also been shown to successfully stabilize topological phases in a harmonic trap.\cite{Wei2012} 
However, another study using the bosonization method reaches a different conclusion, in which the interaction merely shifts the value of Luttinger parameters.\cite{Fidkowski2011}  
They further consider a slightly different scheme where a spin-orbit coupled wire is in contact with a $s$-wave superconductor wire with a power-law decaying order.
The main conclusion of their study is that this model with a single spin-orbit coupled wire does not exhibit Majorana degeneracy and the Majorana zero modes can be obtained only with more than two wires.

One recent work also studied a similar system with a longitudinal magnetic field using the Bethe ansatz exact solution and the conformal field theory.\cite{Zvyagin2013} They obtained the critical exponents of the superfluidity and density waves over a wide range of SOC and found that the superfulid correlation always has a smallest exponent. Their results also show the system has Flude-Ferrell-Larkin-Ovchinnikov(FFLO) type instability with a spin-orbit interaction dependent wave vector which may develop into the FFLO phase when an array of such chains coupled together as in the experiment setup.  

Motivated by these results we apply the unbiased numerical methods to solve the full interacting model in one dimension over a wide range of interaction strength. We also study the dominant correlations with the bosonization method. Our numerical results are obtained using the iTEBD\cite{Vidal2007} and the DMRG\cite{White1992,White1993,Schollwock2005}
methods. The iTEBD method is convenient for studying the phase diagram in the thermodynamic
limit. We keep a virtual dimension up to 1000 and apply the imaginary evolution to reach the ground state. 
On the other hand the DMRG method is known to be accurate for finite size systems and allow us to determine spin excitation gaps in the case without the external magnetic field.
We apply the DMRG method for systems up to 200 sites. 
The $U(1)$ symmetry for particle number conservation is considered. 
In the DMRG method we keep $m=300$ virtual states and apply seven sweeps. The discarded weight in the eigenvalue of the reduced density matrix with this set of parameters is on the order of $10^{-6}$ in the worst case.
 
Our main finding is that two different superconducting phases are stablized besides insulating phases found in the strong field or high chemical potential region.
Both superconducting phases have a zero momentum triplet pairing order which has the same critical exponent as that of the singlet pairing order.
Finite momentum pairing terms as those found in the FFLO phase of the spin imbalanced system without SOC are not dominant in both superconducting phases.
We also investigate the phase space which was suggested to be a topological phase with Majorana fermions in the mean field study but no clear evidence of the Majorana fermion is found in our numerical results.

The rest of this work is organized as following: in Sec. \ref{sect:model} we introduce the model of interest. The phase diagram of the non-interacting system is shown for comparison with that of the interacting one. In Sec.  \ref{sect:bosonization} we follow an earlier work\cite{Gangadharaiah2008} to derive the bosonization Hamiltonian and give the expression of the important order parameters. The phase diagram of the interacting system is given in Sec. \ref{sect:phase}. We also show correlation functions of the different phases and compare their long range behavior with the prediction from the result in Sec. \ref{sect:bosonization}. Finally, we conclude our study in Sec. \ref{sect:conclusion}.

\section{model}
\label{sect:model}

We considered a spin-orbit coupled fermionic atoms in a transverse magnetic field
confined in a one-dimensional optical lattice. The Hamiltonian reads
\begin{align} \label{eq:mainH}
H=&-t\sum_{\langle i,j\rangle,\sigma}(c^{\dag}_{i,\sigma}c_{j,\sigma}+h.c) \\ \nonumber
&-\alpha \sum_{\langle i,j\rangle}[i(c^{\dag}_{i,\uparrow}c_{j,\uparrow}-c^{\dag}_{i,\downarrow}c_{j,\downarrow})+h.c.)] \\ \nonumber
&-h\sum_i (c^{\dag}_{i,\uparrow}c_{i,\downarrow}+c^{\dag}_{i,\downarrow}c_{i,\uparrow})\\ \nonumber
&+U\sum_i n_{i,\uparrow}n_{i,\downarrow}-\mu \sum_i n_{i},
\end{align}%
where $t$ is the hopping strength, $\alpha$ is the Rashba SOC strength\cite{Bychkov1984}, $h$ is magnetic field in the $x$-direction, $\mu$ is the chemical potential and $n_i=c^\dag_ic_i$ is the density operator. In this study we take $t=\alpha=1$ as our energy unit and investigate the $U<0$(attractive) case.

\begin{figure}[htb!]
\includegraphics[width=8cm]{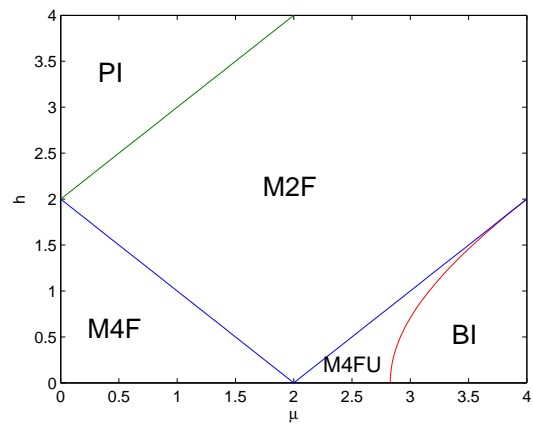}
\caption{The phase diagram of the non-interacting system in the $\mu$-$h$ plane. PI and BI stand for
a polarized insulator and a band insulator respectively. M4F, M2F, and M4FU are
metal phases with four Fermi points, two Fermi points, and all four Fermi points
in the upper bands.} \label{fignonInt}
\end{figure}

In the absence of interaction Eq.~\ref{eq:mainH} can be diagonalized to obtain the energy
eigenvalues
\begin{align}
\epsilon_{\pm}=-2t\cos(k)\pm\sqrt{h^2+4\alpha^2\sin^2(k)},
\end{align}
and the momentum $k$ dependent eigenspinors,
\begin{align}
\vert\chi_{+}(k)\rangle=
\begin{bmatrix}
\sin\frac{\gamma(k)}{2}& \\
\cos\frac{\gamma(k)}{2}& 
\end{bmatrix}
\end{align}
 and
\begin{align}
\vert\chi_{-}(k)\rangle=
\begin{bmatrix}
\cos\frac{\gamma(k)}{2}& \\
-\sin\frac{\gamma(k)}{2}& 
\end{bmatrix}
\end{align}
, where $\gamma(k)$ is defined as
\begin{align}
\gamma(k)=\arctan\frac{h}{2\alpha \sin(k)}.
\label{eq:gamma}
\end{align}
The spin direction of the eigenspinor has a dependence on $k$.
Away from $k=0$, the eigenspinor $\vert \chi_+(k)\rangle$ has a spin component largely aligned 
in the $+z$ direction at $k<0$ and in the $-z$ direction at $k>0$ in the small $h$ 
case while the spin of the eigenspinor $\vert \chi_-(k)\rangle$ aligned in the opposite 
direction. The nonzero spin overlap between these two bands at the same energy as
a consequence of the interplay between SOC and transverse magnetic
field introduces triplet spin pairing at zero momentum as we turn on the attractive interaction.

The momentum at Fermi level in the upper band and lower band is denoted by $k_+$ and $k_-$ respectively. We also define
 \begin{align}
2k_F=k_++k_-,
\end{align}
 \begin{align}
\delta k_F=\vert k_+-k_-\vert,
\end{align}
and 
\begin{align}
v_F=2t\sin(k_F)-2\alpha\cos(k_F).
\end{align}

In Fig.~\ref{fignonInt} we show the phase diagram of the noninteracting system. It is known that the SOC
term lifts the spin degeneracy and shifts the band energy minimum
to a finite momentum. The transverse magnetic field opens up a gap at the crossing of the two bands at zero momentum. Therefore, depending on the chemical potential $\mu$(or filling number $n$)
the noninteracting system can go through several phase transitions. 
We focus on the phase above half-filling since the physics below half-filling is similar due to the particle-hole symmetry.
At half filling $\mu=0$ and $h<2$ we have a phase with four Fermi points, two in 
the lower band and two in the upper band. As $\mu$ increases such that the Fermi 
surface sits inside the $k=\pi$ gap the system enters a phase with two Fermi points 
in the upper band. Further increasing the filling leads to a phase with four Fermi 
points all in the upper band. When $\mu$ goes beyond the maximum of the upper 
band all bands are occupied we thus have a band insulator(BI). In the upper 
left corner of the phase diagram $\mu$ lies in between the gap opened by a strong magnetic 
field, which leads to a fully occupied lower band and an empty upper band. In this case we have an insulator at half-filling due to the strong magnetic
field and the system has a high magnetization. We will denote it as a polarized insulator(PI). 

In the next section we discuss the interaction effect on the system. The case
in which there are two Fermi points has been discussed with the bosonization 
method before.\cite{Fidkowski2011} Here we focus on the filling with four Fermi 
points. With repulsive interaction a bosonization study suggests a spin density 
order in the direction of the spin-orbit axis.\cite{Gangadharaiah2008} Following 
their work we rederive the expression of correlation functions in 
section~\ref{sect:bosonization} and compare with our numerical results in 
section~\ref{sect:phase} for the attractive interaction.

\section{Bosonization prediction}
\label{sect:bosonization}

We study the interacting system by first linearized the spectrum near the 
Fermi points, $\pm k_\pm$. We focus on the case when both upper and lower
band are partially occupied. The case at high filling number where the lower
band is fully occupied and all four Fermi points are in the upper band is similar.
Due to the difference of spin components in these two bands the dominant
momentum component in the correlation function changes but the physical
picture remains the same.

The fermionic operator can be written in terms of
left-going particles and right-going particles of $\nu=\pm$ bands.
\cite{Gangadharaiah2008}
\begin{align}
\Psi_\sigma(x)=\sum_{\nu=\pm}\langle\chi_\nu(k_\nu)\vert\sigma\rangle e^{ik_\nu x}R_\nu
+\langle\chi_\nu(-k_\nu)\vert\sigma\rangle e^{-ik_\nu x}L_\nu
\end{align}
We then follow the convention used in the Ref.~\bibpunct{}{}{;}{n}{,}{,}\cite{Gangadharaiah2008}\bibpunct{}{}{;}{s}{,}{,}
to bosonize the fermionic operators with
\begin{align}
R_{\pm}=\frac{\eta_\pm}{\sqrt{2\pi a_0}}e^{i\sqrt{4\pi}\phi_{R\pm}},~
L_{\pm}=\frac{\eta_\pm}{\sqrt{2\pi a_0}}e^{-i\sqrt{4\pi}\phi_{L\pm}},
\end{align}
where $a_0\sim k_F^{-1}$ is the short-distance cutoff and $\eta_\pm$ are the
Klein factors with $\eta_+\eta_-=i$. The chiral $\phi_{R/L\nu}$ are expressed in
terms of $\phi_\nu$ and its dual $\theta_\nu$ as
\begin{align}
\phi_{R\nu}=\frac{\phi_\nu-\theta_\nu}{2},~\phi_{L\nu}=\frac{\phi_\nu+\theta_\nu}{2},
\end{align}
and the chiral densities are written in terms of
\begin{align}
R^\dag_\nu R_\nu=&\frac{\partial_x\phi_{R\nu}}{\sqrt{\pi}}=\frac{\partial_x(\phi_{\nu}
-\theta_\nu)}{\sqrt{4\pi}}, \nonumber \\
L^\dag_\nu L_\nu=&\frac{\partial_x\phi_{L\nu}}{\sqrt{\pi}}=\frac{\partial_x(\phi_{\nu}
+\theta_\nu)}{\sqrt{4\pi}}.
\end{align}

The Hamiltonian can then be written in terms of charge $\phi_\rho, \theta_\rho$ and
pseudo-spin $\phi_\sigma, \theta_\sigma$ modes defined as
\begin{align}
\phi_{\rho}=\frac{\phi_--\phi_+}{\sqrt{2}},~&\phi_{\sigma}=\frac{\phi_-+\phi_+}{\sqrt{2}}, 
\nonumber \\
\theta_{\rho}=\frac{\theta_--\theta_+}{\sqrt{2}},~&\theta_{\sigma}=\frac{\theta_-+\theta_+}{\sqrt{2}}.
\end{align}
The quadratic part of the Hamiltonian reads
\begin{align}
H_{\rho/\sigma}=\frac{1}{2}\int dx\left[u_{\rho/\sigma} K_{\rho/\sigma} 
(\partial_x\theta_{\rho/\sigma})^2+\frac{u_{\rho/\sigma}}{K_{\rho/\sigma}}
(\partial_x\phi_{\rho/\sigma})^2\right].
\end{align}
Following Ref.~\bibpunct{}{}{;}{n}{,}{,}\cite{Gangadharaiah2008}\bibpunct{}{}{;}{s}{,}{,} the only relevant perturbation term is
\begin{align}
H^C_\sigma=\frac{U\cos^2(\frac{\gamma_++\gamma_-}{2})}{2(\pi a_0)^2}\int dx
\cos(\sqrt{8\pi}\theta_\sigma),
\label{eq:Cooper}
\end{align}
which comes from the process of a pair of oppositely moving particles
in the same subband $\nu$ scattering in to the other subband. 
The $\cos^2(\frac{\gamma_++\gamma_-}{2})$ factor describes the spin
overlap in the process. Since this term is the only relevant perturbation it determines the phase
of the system as we can see in the following discussion.

The Luttinger liquid parameters, $K_{\rho/\sigma}$ and velocities $u_{\rho/\sigma}$
are obtained in the first order of $U/v_F$ as
\begin{align} \label{eq:LTparameters}
u_\rho\approx &~v_F\left[1+\frac{U}{\pi v_F}\right], \nonumber \\
u_\sigma\approx &~v_F\left[1-\frac{U\cos^2(\gamma_F)}{2\pi v_F}\right], \nonumber \\
K_\rho\approx &~1-\frac{U}{2\pi v_F}(1+\frac{\sin^2(\gamma_F)}{2}), \nonumber \\
K_\sigma\approx &~1-\frac{U}{2\pi v_F}(\cos^2(\gamma_F)-\frac{\sin^2(\gamma_F)}{2}),
\end{align}
where $\gamma_F\equiv\gamma(k_F)$ and $\gamma(k_\pm)\approx\gamma_F$ is 
assumed.

\begin{figure}[htb!]
\begin{center}
\includegraphics[width=8cm]{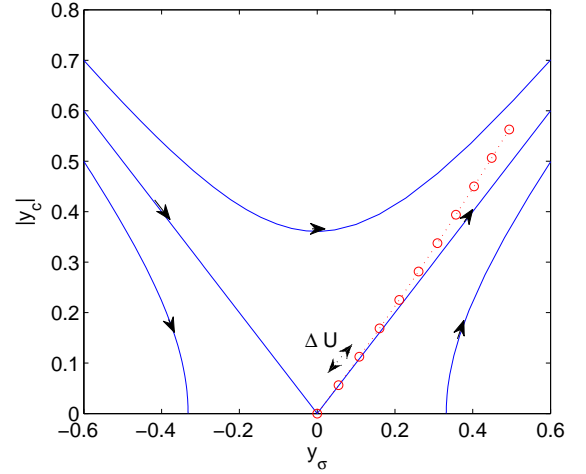}
\caption{(Color online) Blue solid lines are RG flows solved from Eq.~\ref{eq:RG}.
Red dots are initial values of $y_C$ and $y_\sigma$ at $h=0.2$, $n=1.1$, and 
$U=0\sim-5$ with a step $\Delta U=0.5$.} \label{fig:RGflow}
\end{center}
\end{figure}

For the purpose of renormalization group(RG) analysis we define the initial value of 
interaction parameters as
\begin{align}
y_\sigma(l=0)&=2(K_\sigma-1)\nonumber\\&=-\frac{U}{\pi v_F}(\cos^2(\gamma_F)-\frac{\sin^2(\gamma_F)}{2}), \nonumber \\
y_C(l=0)&=\frac{U\cos^2(\frac{\gamma_++\gamma_-}{2})}{\pi u_\sigma}.
\end{align}
The RG equations obtained is the standard KT equations\cite{Gogolin1999,Giamarchi2004}:
\begin{align}
\frac{dy_\sigma}{dl}=y^2_C,\;\frac{dy_C}{dl}=y_\sigma y_C.
\label{eq:RG}
\end{align}
Depending on the value of the interaction strength $U$ and $\gamma_F$ the RG flow
can go to strong coupling or a Luttinger liquid(LL) phase. In Fig.~\ref{fig:RGflow} we plot the RG flow 
of Eq.~\ref{eq:RG} and initial values of $y_\sigma$ and $y_C$ for several interaction 
strength $U$ at a filling number $n=1.1$ and $h=0.2$.
We observe that for these parameters and an attractive interaction the RG flow goes to the strong coupling limit. In fact for our choice of $\alpha=1$ the $y_\sigma$ value is always positive in the M4F region as long as $U<0$, which means the system will always flow into the strong coupling limit as we can see in Fig.~\ref{fig:RGflow}.
Although for $\alpha<h/2\sqrt{2}$, $y_\sigma$ will have a negative value, we found in these cases $|y_c|$ is always larger than $|y_\sigma|$. Therefore, we conclude that there is no transition to the LL phase in the M4F region when the interaction is turned on.

In order to determine the dominant order of the system we write down here the
expression of several correlations in their bosonization forms. The
coefficients of different momentum components depend on the value of $\gamma$ in Eq.~\ref{eq:gamma}. 
Here we approximate
$\gamma(k_\pm)\simeq\gamma(k_F)$ and list only the momentum component with the largest coefficients for each order parameter in the low $h$ case where
\begin{align}
\lim_{h\to0}\gamma\simeq0.
\end{align}
We note that for general values of $h$, the system can have slow decaying correlations with several different momentum components. This complicates the behaviour of the correlations. For the purpose of comparing with our numerical results in the next section we will only focus on the small field case here.
The finite momentum component of spin density($O^{SDW}$) and charge density($O^{CDW}$) terms with the largest coefficients are
\begin{align}
O^{SDW_x}(x)\simeq~&\sin(\sqrt{2\pi}(\phi_\rho\mp\phi_\sigma)+2k_\pm x), \\
O^{SDW_y}(x)\simeq~&\cos(\sqrt{2\pi}(\phi_\rho\mp\phi_\sigma)+2k_\pm x), \\
O^{SDW_z}(x)\simeq~&\cos(\sqrt{2\pi}\phi_\rho+2k_Fx)\sin(\sqrt{2\pi}\theta_\sigma), \\
O^{CDW}(x)\simeq~&\sin(\sqrt{2\pi}\phi_\rho+2k_Fx)\cos(\sqrt{2\pi}\theta_\sigma).
\label{eq:CDW}
\end{align}
The singlet($\Delta_s$) and triplet($\Delta_T$) pairing both have a zero momentum 
term $e^{i\sqrt{2\pi}(\theta_\rho\pm\theta_\sigma)}$ and finite momentum terms
\begin{align}
\Delta_s(x)\simeq~&e^{i\sqrt{2\pi}\theta_\rho}\cos(\sqrt{2\pi}\phi_\rho+2k_Fx), \\
\label{eq:PT}
\Delta_T(x)\simeq~&e^{i\sqrt{2\pi}\theta_\rho}\sin(\sqrt{2\pi}\phi_\sigma+\delta k_Fx).
\end{align}

When the RG flow goes to the strong coupling limit in order to minimize the energy, the $\theta_\sigma$ field in Eq.~\ref{eq:Cooper} will take the semiclassical minima $\theta_\sigma=(m+\frac{1}{2})\sqrt{\pi/2}$ or $\theta_\sigma=m\sqrt{\pi/2}(m\in\mathbb{Z})$ for positive or negative $U$ respectively.
With attractive interaction we find from the above expression that the system will have non-oscillating singlet and triplet pairing order with the same smallest exponent $1/K_\rho$. The finite momentum terms in the singlet pairing signal a FFLO type instability.
However, these terms always have a larger exponent when the system goes to the strong coupling limit. This is quite different from the case without SOC, where FFLO is found in a considerable part of the phase diagram.\cite{Heidrich-Meisner2010}
The sub-dominant order in the system is the $2k_F$ charge-density wave with an exponent $K_\rho$.
We will compare these prediction with our numerical results in the next section.

\section{Phase diagram}
\label{sect:phase}

\begin{figure}[htb!]
\begin{center}
\includegraphics[width=9cm]{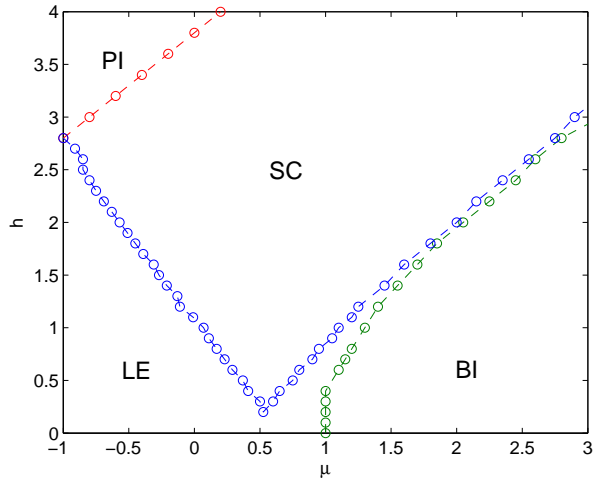}
\caption{(Color online) The phase diagram of the interacting system with $U=-2$.
SC stands for a superconducting phase. LE is a pseudo-spin gap superconducting phase.(see context)} \label{fig1}
\end{center}
\end{figure}

Our results of the attractively interacting system can be summarized in Fig.~\ref{fig1}. 
We map out the phase diagram in the $\mu$-$h$ plane with attractive interaction $U=-2$. 
Due to the particle-hole symmetry of the Hamiltonian the system is at exact half-filling when $\mu=-U/2$. We will focus on the phase diagram above half-filling. 
The phase diagram less than half-filling can be obtained by performing a particle-hole transformation.

Two insulating phases similar to those in the non-interacting system are located at two corners. 
We find a polarized insulator(PI) in the upper-left corner of the phase diagram. 
Due to the strong transverse field, a gap opened between the upper and the lower band. 
The ground state on a single site is close to $(\vert\uparrow\rangle+\vert\downarrow\rangle)/\sqrt{2}$ with filling number 1.
On the lower-right corner we have a band insulator(BI) with both bands fully occupied. 

In the middle region a fully occupied lower band and a partially filled upper band developed into a Luttinger liquid phase when we turn on the interaction. Since this phase has a dominant correlation with superconductor orders we will denote it as a SC phase.

In the lower-left corner we find a phase transition from the SC phase to a another superconducting phase.
We identify this phase as the gapped superconducting phase discussed in the earlier section. We will call it as the Luther-Emery phase(LE) due to the similarity of the gapped nature. This phase is developed from the non-interacting M4F region. The M4FU region in the non-interacting system is now connected with the LE phase at small $h$ when a finite interaction $U$ is turned on. 
The BI phase at finite $h$ extends to the $h=0$ line while another superconducting phase at $h=0$ separates itself from the LE phase. We will show that this phase has a well-defined spin gap and its correlations behave differently from those in the LE phase.

\begin{figure}[htb!]
\begin{center}
\includegraphics[width=9cm]{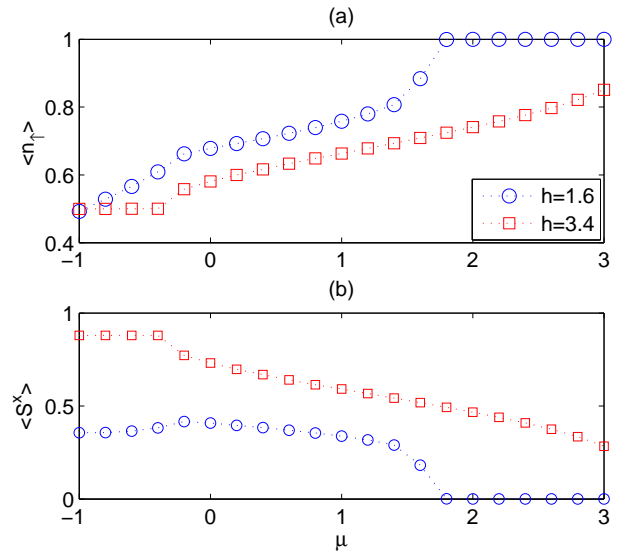}
\caption{(Color online) Order parameters $\langle n_\uparrow\rangle$ and
$\langle S^x\rangle$ as functions of $\mu$ at $h=1.6$ (blue dot) and $h=3.4$ (red square). The phase
boundary is determined from the discontinuity of slopes.} \label{fig1_OP}
\end{center}
\end{figure}

We determined the phase boundary by calculating the order parameters $\langle S^x
\rangle$ and $\langle n\rangle$ as a function of the chemical potential $\mu$ at different field strength $h$ with the iTEBD method. 
In Fig.~\ref{fig1_OP} we show the results at $h=1.6$ and $h=3.4$ as an example. At $h=1.6$ we observe a discontinuity of the slopes of both order parameters at $\mu=-0.2$, 1.4, and 1.8, which signals phase 
transitions at these points. The phase beyond $\mu=1.8$ has zero magnetization and
a filling number 2. Thus it is a BI. Other three phases show quite different behavior of
compressibility and magnetization as $\mu$ changes and do not have simple features to be identified. 
Similarly we can identify a single transition at $\mu=-0.4$ in the $h=3.4$ line. 
The magnetization saturates to a fixed value when $\mu\leq-0.4$ and the filling number is 1, which suggest it is the PI phase as that in the non-interacting system.

\subsection{LE phase}

In order to understand those phases other than PI and BI we compute several correlation functions to characterize them. We define the density-density correlation functions,
\begin{align}
N_{ij}=\langle n_in_j\rangle-\langle n_i\rangle\langle n_j\rangle,
\end{align}
spin $\beta=x, y, z$ component correlation function,
\begin{align}
S^\beta_{ij}=\langle S^\beta_iS^\beta_j\rangle-\langle S^\beta_i\rangle\langle S^\beta_j\rangle,
\end{align}
the pairing correlation function
\begin{align}
P_{s/T,ij}=\langle \Delta^\dag_{s/T,i}\Delta_{s/T,j}\rangle,
\end{align}
where
\begin{align}
\Delta^\dag_{s,i}=c^\dag_{i,\uparrow}c^\dag_{i,\downarrow}
\end{align}
for on-site singlet pairing, and
\begin{align}
\Delta^\dag_{T,i}=c^\dag_{i,\uparrow}c^\dag_{i+1,\uparrow}
\end{align}
for triplet pairing.
The corresponding structure factors are Fourier transformation of the above
correlation function,
\begin{align}
X(k)=\frac{1}{L}\sum^L_{i,j=1}e^{\imath k(x_i-x_j)}X_{ij},
\end{align}
where $L$ is the system size.

\begin{figure}[htb!]
\includegraphics[width=9cm]{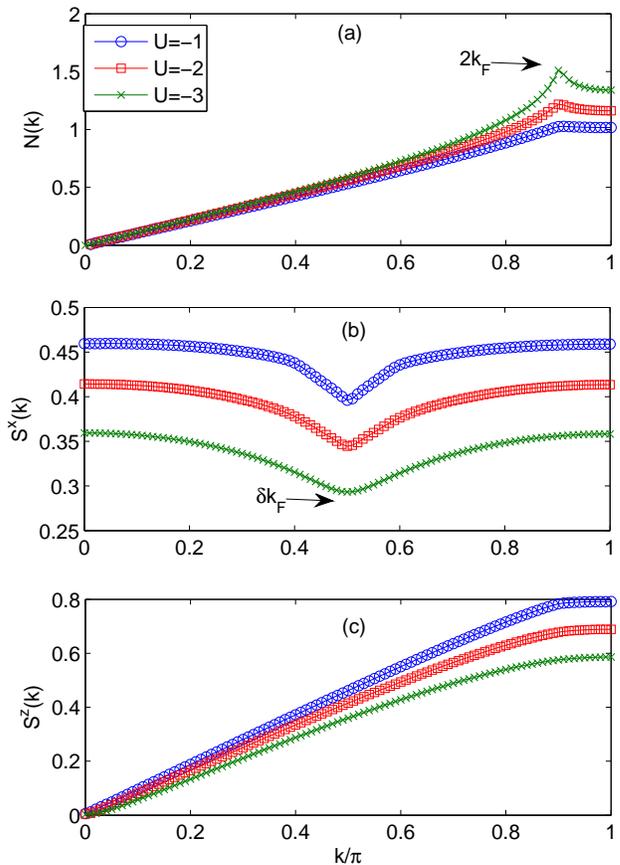}
\caption{(Color online) Structure factors of (a) density $N(k)$, (b) spin density
$S^x(k)$, and (c) spin density $S^z(k)$ at $U=-1$(blue dot), -2(red square),
-3(green cross), $h=0.2$ and $n=1.1$.} \label{fig3}
\end{figure}

\begin{figure}[htb!]
\includegraphics[width=9cm]{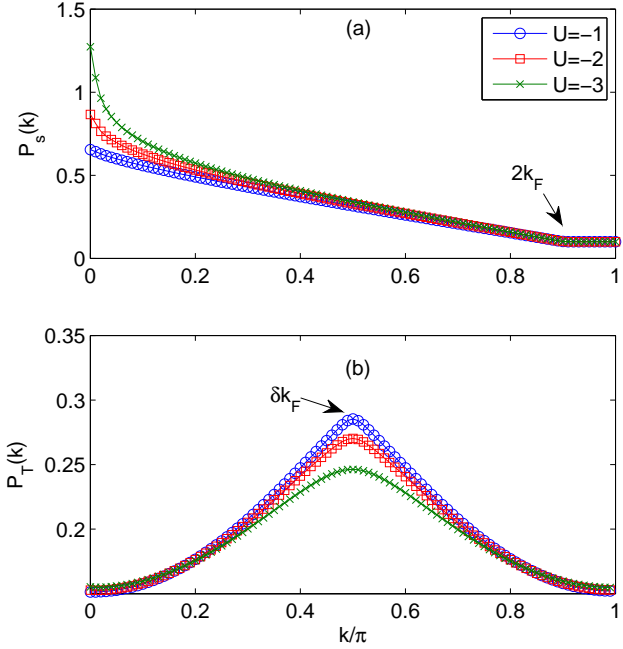}
\caption{(Color online) Structure factors of (a) singlet pairing $P_s(k)$ and (b) triplet
pairing $P_T(k)$ at $U=-1$(blue dot), -2(red square), -3(green cross), $h=0.2$, and
$n=1.1$.} \label{fig4}
\end{figure}

In Fig.~\ref{fig3} we plot the density wave and the spin density structure factors at $h=0.2$, $n=1.1$ with $U=-1$, -2, and -3 for $L=200$, which belongs to the phase in the bottom-left corner in Fig.~\ref{fig1}. 
We observe in Fig.~\ref{fig3}(a) a cusp develops at $2k_F$ momentum of density-density structure factor, which signals a $2k_F$ charge-density wave as we increase the attractive interaction to $U=-3$. $S^x$ density structure factor has a kink at $2k_\pm$ and a dip at $\delta k_F$ momentum. 
The kinks at $2k_\pm$ are smoothed as $U$ increases as shown in Fig.~\ref{fig3}(b). 
Meanwhile, we find in Fig.~\ref{fig3}(c) the $2k_F$ structure factor of $S^z$ density are suppressed in all three interaction strength.
From these results we can infer that the dominant order in the density wave part is the charge density one.

In Fig.~\ref{fig4}(a) we observe the singlet pairing structure factor develops a cusp at $k=0$ as we increase the strength of attractive interaction, which suggests a strong singlet pairing order at large $U$. 
Together with the $2k_F$ density wave order this corresponds to a strong coupling Luther-Emery(LE) phase discussed in \ref{sect:bosonization}. 
We will further study their real space correlations below. 
A kink at $2k_F$ momentum is also observed in the singlet pairing, which also agree with the bosonization result.
In Fig.~\ref{fig4}(b) we can find a bump at $\delta k_F$ momentum in the triplet pairing structure factors at all interaction strength. 
However, the supposed dominant $k=0$ component predicted by the bonsonization analysis does not have a prominent peak.
This can be understood by a smaller coefficient of the $k=0$ component as we will see when the correlations are plotted in the real space.

\begin{figure}[htb!]
\includegraphics[width=9cm]{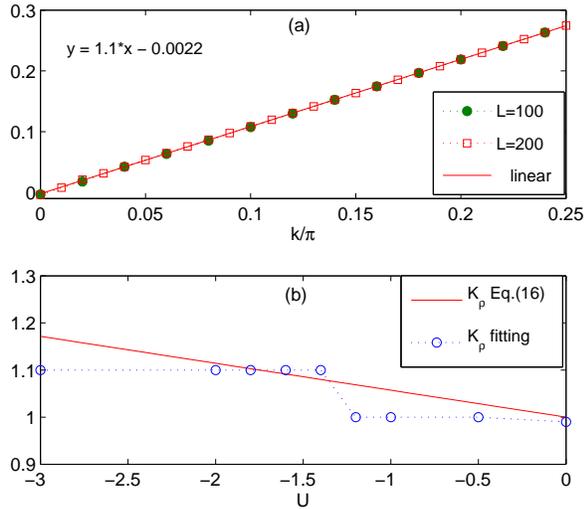}
\caption{(Color online) (a) Structure factor $N(k)$ of $L=100$ (blue dot) and $L=200$ (red
square) at small $k$. The linear fitting result is shown as a solid line. (b) Luttinger 
parameter $K_\rho$ obtained from Eq.~\ref{eq:LTparameters}(solid line) and
from linear fitting results(blue dot) as a function of $U$.} \label{fig:Krho}
\end{figure}

The bosonization method suggests with any nonzero attractive interaction the system will flow into strong coupling limit. 
The exponents of dominant correlation functions can
then be determined by a single Luttinger parameters $K_\rho$. In order to compare this 
conclusion with our numerical study we first extract the Luttinger parameter $K_\rho$ from
the density structure factor. It can be shown that in the $k\to0$ limit the density structure factor
reads\cite{Clay1999,Moreno2011}
\begin{align}\label{eq:fitkrho}
N(k)\to K_\rho k \pi.
\end{align}
We shows in Fig.~\ref{fig:Krho}(a) the linear fitting result of the density structure 
factor near $k=0$ at $U=-2$, $h=0.2$, and $n=1.1$ with $L=100$ and $200$.
We find that finite size effects do not significantly change the fitting results. 
The $K_\rho$ values extract from these two system sizes agree with each other.
Thus, we only show the results for $L=200$ in our data below.
In Fig.~\ref{fig:Krho}(b) we compare the $K_\rho$ values from numerical results 
with those calculated in Eq.~\ref{eq:LTparameters} and find that they match reasonably
well within the interaction strength regime we studied.

We now turn to the real space correlations. 
It is known that the DMRG calculation of real space correlations may suffer from both finite size effects and Friedel oscillation from fixed boundary condition\cite{Bedurftig1998}. 
One solution widely used is to average over correlations in only the center part of the system, which is taken to be far from the boundary.
On the other hand the iTEBD method exploit the translational symmetry of the system and can be used to calculated correlations without worrying about
the boundary and finite size effects. 
In a recent work it has been used to study the correlations of the spinless fermion system, in which good agreement with the bosonization method is obtained\cite{Karrasch2012} In the following we will mainly present the correlations from the iTEBD results. The DMRG results will serve
as a comparison in some cases.

\begin{figure}[htb!]
\includegraphics[width=9cm]{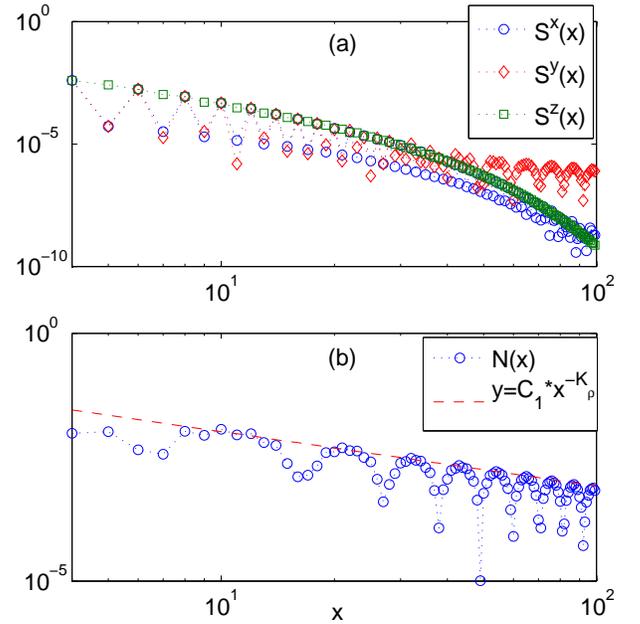}
\caption{(Color online) Real space correlation functions (a) $S^x(x)$ (blue dot), $S^y(x)$ (red diamond), $S^z(x)$ (green square) and (b) $N(x)$ in log-log scale at $U=-3$, $h=0.2$, and $n=1.1$. The red dashed line in (b) is a polynomial fitting function with the slope 
determined from Eq.~\ref{eq:fitkrho}.} \label{fig:CSDW}
\end{figure}

In Fig.~\ref{fig:CSDW}(a) we plot the spin density correlation functions in log-log scale at $U=-3$, $h=0.2$, and $n=1.1$. We find that both $S^x$ and $S^y$ spin components have a correlation which decays exponentially in the short range but has a polynomially-decaying tail while the $S^z$ correlation decays exponentially. 
Furthermore, both $S^x$ and $S^y$ modulated with the same periodicity at short distance and has a different modulation similar to that of the density correlation at long distance.

This behavior can be understood from the bosonization results. 
Since $\theta_\sigma$ takes the value $m\sqrt{\pi/2}$ in the strong coupling limit, the $S^z(x)$ density wave, which has a bosonization form $\sin(\sqrt{2\pi}\theta_\sigma)$, will have a zero expectation value. 
This leads to a correlation which decays exponentially. 
The fluctuation of the conjugate variable $\phi_\sigma$ also leads to a zero expectation value of $S^x(x), S^y(x)\sim\sin(\sqrt{2\pi}\phi_\sigma)$, which explains the short-range behavior of them.
Besides the short-ranged term, both of them have a $2k_F$ term with smaller coefficients described by
\begin{align}
O^{SDW_x}(x)\simeq~\sin\frac{\gamma_+-\gamma_-}{2}\sin(\sqrt{2\pi}\phi_\rho+2k_Fx)\cos(\sqrt{2\pi}\theta_\sigma), \nonumber\\
O^{SDW_y}(x)\simeq~\sin\frac{\gamma_++\gamma_-}{2}\cos(\sqrt{2\pi}\phi_\rho+2k_Fx)\cos(
\sqrt{2\pi}\theta_\sigma).
\label{eq:SxSy}
\end{align}
In the strong coupling limit $\cos(\sqrt{2\pi})\theta_\sigma$ takes a finite value and both terms decay polynomially. 
However, since $\gamma\to0$, these terms show up only after a certain distance where the exponentially-decaying terms vanish.
The bosonization result also explains different modulation periods at the short range and the long range.
We also check this behavior at stronger interaction $U=-8$ (not shown), where we find a similar result. Due to the shorter correlation length at $U=-8$, the $2k_F$ terms start to dominate at a shorter distance.

\begin{figure}[htb!]
\includegraphics[width=9cm]{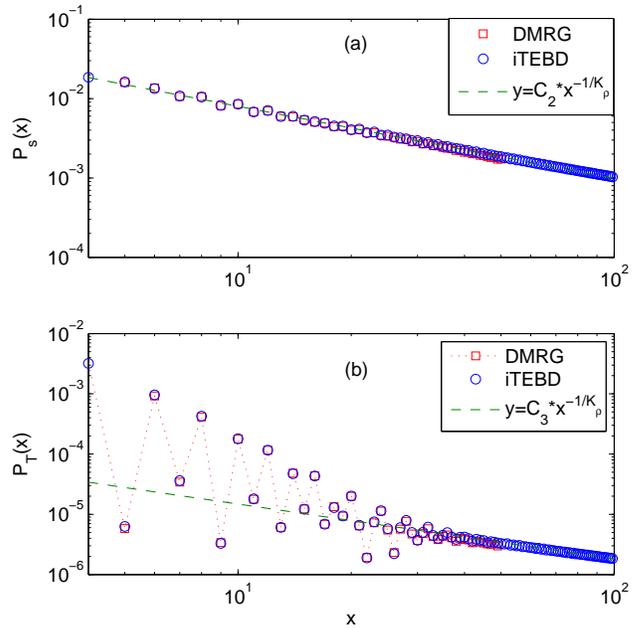}
\caption{(Color online) Real space correlation functions (a) $P_s(x)$ and (b) $P_T(x)$ determined by the DMRG (red square) or the iTEBD (blue dot) method
in log-log scale with the same parameter as in Fig.~\ref{fig:CSDW}. 
The green dashed lines in (a) and (b) show the
asymptotic behavior determined by the bosonization method.} \label{fig:SSTS2}
\end{figure}

The charge density correlations $N(x)$ and singlet pairing correlation $P_s(x)$ 
are now given by
\begin{align}
\label{eq:Nxreal}
N(x)&\sim\cos(2k_Fx)x^{-K_\rho} \\
\label{eq:Psreal}
P_s(x)&\sim x^{-1/K_\rho}
\end{align}
since the $\theta_\sigma$ in Eq.~\ref{eq:CDW}
takes it semiclassical value in the LE phase. In Fig.~\ref{fig:CSDW}(b) and~\ref{fig:SSTS2}
we show the correlation functions and their asymptotic forms.
The $K_\rho=1.1$ used in the asymptotic forms is obtained from Eq.~\ref{eq:fitkrho}. 
The $C_1$ constant is determined by fitting the maximum values of the correlations in the $2k_F$ oscillations.
We can clearly see that the power law behavior is well-described by Eq.~\ref{eq:Nxreal}.
Furthermore, the same periodicity of the charge density correlation and the long range term of the $S^x$ and $S^y$ correlation seen in Eq.~\ref{eq:Nxreal} and Eq.~\ref{eq:SxSy} also agrees with our numerical results.
Similarly, the non-oscillating component of the singlet paring correlation agrees nicely with Eq.~\ref{eq:Psreal} as we show in Fig.~\ref{fig:SSTS2}(a).

In Fig.\ref{fig:SSTS2}(b) we plot the modulus of the real part of the triplet correlation.
We have shown earlier the non-oscillating part of the triplet correlation possesses exactly the same exponent as that of the singlet correlation from the bosonization results. 
Besides, it also has a $\delta k_F$ momentum term with the largest coefficient at small $h$ field, which can readily be seen in Fig.~\ref{fig4}(b). Due to the fluctuation
of $\phi_\sigma$ in the LE phase this term decays exponentially.
This explains what we observe in Fig.\ref{fig:SSTS2}(b). The triplet correlation has a oscillating part which decays rather faster and vanishes after 30 sites. 
In the long distance the $k=0$ component dominates. 
We find the long range behavior can be well described by the power law function with the exponent $-1/K_\rho$ as we show by the green dashed line in which
the $C_3$ constant is obtained by fitting only $x>40$ part of the correlation. 

We also demonstrate similar results computed with the DMRG method.
For the DMRG calculation we keep $m=300$ states and work with a system size $L=200$. 
In order to reduce the boundary effect only the correlations in the central 100 
sites are averaged. In Fig.~\ref{fig:SSTS2}(a) and (b)
we plot the singlet and the triplet correlation calculated from both the iTEBD and the DMRG method. 
As we can see two results agree well in the short distance while the DMRG data decay a bit faster in the long range part of the singlet correlation. 
We suspect this may due to the limited virtual dimension $m$ we used here and the finite size effect. 
Since for the iTEBD method with the same computational resource a larger number of $m$ can be used it gives better results in the long distance. Typically, we keep $m=1000$(300) states in the iTEBD calculation of the singlet(triplet) correlation.

Although we have a fairly good agreement between the bosonization results and the numerical one at a strong interaction strength here, we also find that at a smaller $U$ the power low exponents for the singlet pairing correlation are always more negative than $-1/K_\rho$, which suggests a finite $K_\sigma$ value is required to fit the correlation functions. 
Since no signature of phase transition is found in our study as we change $U$ over a wide range we suspect our numerical methods could not capture the real exponents of the correlation functions because of the finite size effect from the limited size $L$ and the limited virtual dimension $m$.
We note that the correlation length of the system at small interaction could be quite large, which means in order to get the correct exponents both the system size and the inherent correlation length determined by the virtual dimension $m$ should be larger than the correlation length of order parameters. This can only be achieved by keeping more states in the DMRG or the iTEBD.

Our numerical results demonstrate that the LE phase is a superconducting phase with a dominant singlet pairing at $k=0$ momentum. 
Although the pairing at finite momentum does occur it has a larger exponent than that of the zero momentum term, which suggests the FFLO phase found in the spin imbalanced system without the SOC is not the main instability here in the spin-orbit coupled system.
We also showed that the triplet pairing correlation has the same $k=0$ component but the amplitude of the correlations is several orders smaller than that of the singlet one. Hence, the short-ranged behavior of the triplet pairing is dominated by a finite momentum term. 
We further check this conclusion with a stronger interaction $U=-8$.
We find that similar behaviors are seen at the same filling number and external field but the exponentially decaying terms vanishes within tens of sites. 
This can be understood as a shorter correlation length and a larger gap is developed with a stronger interaction.
No further transition is found in the LE phase as we increase the strength up to $U=-8$.

\subsection{Zero field SC phase}

The ground state at the $h=0$ line is a spin gapped superconducting phase. 
At zero field spin components of non-interacting bands do not mix with each other and the spin up and the spin down atoms just have the opposite momentum. 
Once we turns on the interaction the formation of singlet pairs dominates and brings in a spin excitation gap.

\begin{figure}[htb!]
\includegraphics[width=9cm]{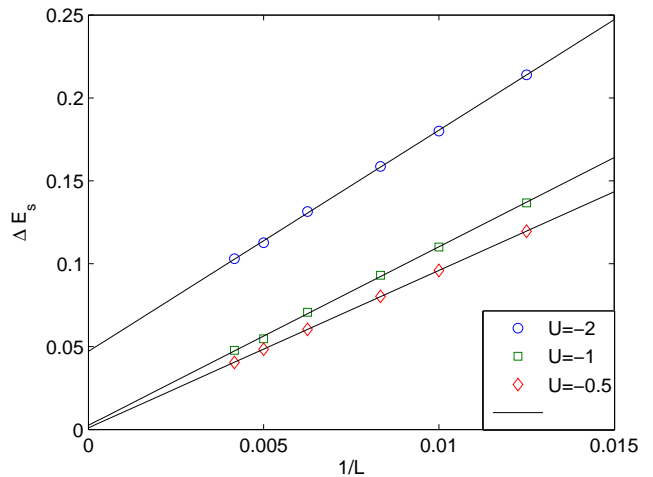}
\caption{(Color online) Spin gaps $\Delta E_s$ at $h=0$, $n\sim1.1$ and
$U=-2$(blue dot), $U=-1$(green square), and $U=-0.5$(red diamond). 
The extrapolations have finite intercept in all cases.} \label{fig:gapvsL}
\end{figure}

To verify the existence of a spin gap at zero field we can measure it with the DMRG method at finite sizes. This can be done by assigning the $S^z_{tot}$ as an additional good quantum number of the system. The spin gap
$\Delta E_s$ is given by the energy difference of ground states with different
$S^z_{tot}$, namely
\begin{align}
\Delta E_s=E_0(N,S^z_{tot}=1)-E_0(N,S^z_{tot}=0),\nonumber
\end{align}
where $E_0(N,S^z_{tot})$ is the ground state energy of given quantum numbers $N$ and $S^z_{tot}$. 
We measure the $\Delta E_s$ at $L=80,\:120,\:160,\:200$ and $240$ and extrapolate them to the thermodynamic limit with a linear function.
Figure~\ref{fig:gapvsL} shows spin gaps as a function of $1/L$ at $h=0,\:n\sim1.1$ and $U=-2,\:-1$, and $-0.5$. 
We can see that spin gaps decrease rather fast as we lower the interaction strength but remain finite in the thermodynamic limit. 
At $U=-0.5$ the $\Delta E_s$ reads $\sim0.001$. 
This conclusion remains at the smallest interaction strength we studied where we have $\Delta E_s=0.00075$ at $U=-0.1$. 

\begin{figure}[htb!]
\includegraphics[width=9cm]{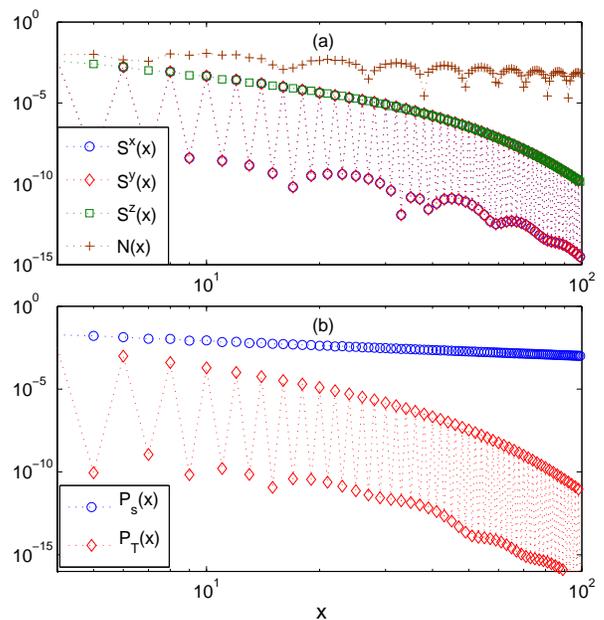}
\caption{(Color online) (a) Real space $S^x(x)$ (blue dot), $S^y(x)$ (red diamond), $S^z(x)$ (green square) and $N(x)$ (brown cross) density correlation functions. (b) $P_s(x)$ (blue dot) and $P_T(x)$ (red cross) correlation functions  at $h=0$, $n\sim1.1$ and $U=-3$. 
} \label{fig:h0SC}
\end{figure}

We could roughly estimates the correlation length $\xi$ from spin gaps data.
The estimation is given by\cite{Gangadharaiah2008}
\begin{align}
\xi\sim\frac{v_F}{\Delta E_s}.\nonumber
\end{align}
The $v_F$ at $h=0$ is roughly at the order of 1, which in turns gives a
correlation length $\xi\sim1000$ at $U=-0.5$. This may justify our conclusion
of the failure to reproduce the correct exponents in the correlation functions
with the DMRG or iTEBD method at small interaction strength.

The spin gap can also be inferred from the long-ranged behavior of the correlation functions. In Fig.~\ref{fig:h0SC}, we plot the spin, charge density, and pairing correlations in the real space with the iTEBD method.
We found in Fig.~\ref{fig:h0SC}(a) that all three spin density correlations decay exponentially. 
Since $U(1)$ symmetry is restored without the external field, $S^x$ and $S^y$ have exactly the same correlations.
The triplet pairing correlation in Fig.~\ref{fig:h0SC}(b) also decays exponentially.
These results agree with the existence of a spin gap from the DMRG calculation.

When comparing correlations with those at $h=0.2$ shown in Fig.~\ref{fig:CSDW} and \ref{fig:SSTS2}, we found the ground state at zero field also has a dominant singlet pairing at $k=0$ and a sub-dominant charge density correlation at $2k_F$. 
Significant difference can be seen at the $S^x$, $S^y$ density and the triplet pairing correlations. 
At a finite field all correlations decay polynomially at the long distance while those at zero field simply have exponentially-decaying correlations and no slow decaying tails are observed up to $100$ sites.
These results remain true at a stronger interaction $U=-8$, in which slow decaying correlations could reveal themselves at a shorter distance due to faster decaying exponential terms if they exist.
We also find that this spin gapped phase transits into the LE phase as long as a small field turns on.
This is quite different from the case without the SOC, in which a spin gapped superconducting phase survives a small finite field when the system has a strong attractive interaction.\cite{Heidrich-Meisner2010}

\subsection{SC phase}

\begin{figure}[htb!]
\includegraphics[width=9cm]{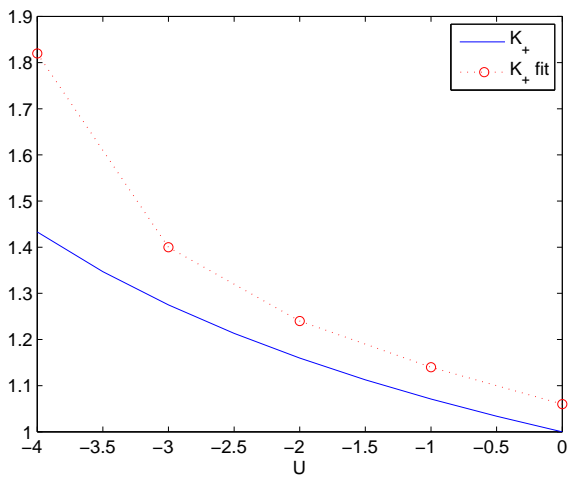}
\caption{(Color online) Luttinger parameter $K_+$ obtained from 
Eq.~\ref{eq:K+}(solid line) and from linear fitting results(red dot) as 
functions of $U$ at $h=1$ and $n=1.5$.} \label{fig:K+}
\end{figure}

At a higher magnetic filed and a larger filling number the system 
enters into another superconducting phase. 
The free system has a fully occupied lower band and two Fermi points in the upper band. As discussed in
Ref.~\bibpunct{}{}{;}{n}{,}{,}\cite{Fidkowski2011}\bibpunct{}{}{;}{s}{,}{,}, interaction effect merely shifts the free Luttinger
parameter $K_+=1$. We find that the dependence of $K_+$ on $U$ reads\cite{
Gangadharaiah2008}
\begin{align} \label{eq:K+}
K_+=\left(\frac{1}{1+\frac{U\cos^2(\gamma_+)}{\pi v_+}}\right)^{\frac{1}{2}},
\end{align}
where $v_+$ is the fermi velocity in the upper band. In Fig.~\ref{fig:K+} we compare 
the dependence of $K_+$ on $U$ from Eq.~\ref{eq:K+} and the numerical results obtained
from the structure factors of the charge density wave at $h=1$ and $n=1.5$.
We find a systematic overestimate from numerical results at $|U|\leq3$. The
relative error is about $6\sim7\%$. With even stronger 
attractive interaction $|U|\geq4$ our numerical data shows a much higher $K_+$ than
those determined by Eq.~\ref{eq:K+} we suspect the bosonization method fails 
in this strong coupling region.

The bosonized spin density and charge density operators have a $k=0$ component
proportional to $\nabla \theta_+$ and $\nabla \phi_+$ respectively. The $2k_+$
momentum component of the charge density operator is
\begin{align}
O^{CDW}(x)\simeq~&\sin(\sqrt{4\pi}\phi_++2k_+ x).
\end{align}
Their real space correlations then read
\begin{align}
S^z(x)&\simeq A_1\frac{1}{x^2}\label{eq:Szplus} \\ 
N(x)&\simeq \frac{K_+}{2(\pi x)^2}+B_2\frac{\cos(2k_+ x)}{x^{2 K_+}}. \label{eq:Nxplus}
\end{align}
Since $K_+>1$ for the attractive interaction the $k=0$ term dominates and both
correlations have a decay exponent -2.

\begin{figure}[t!]
\includegraphics[width=9cm]{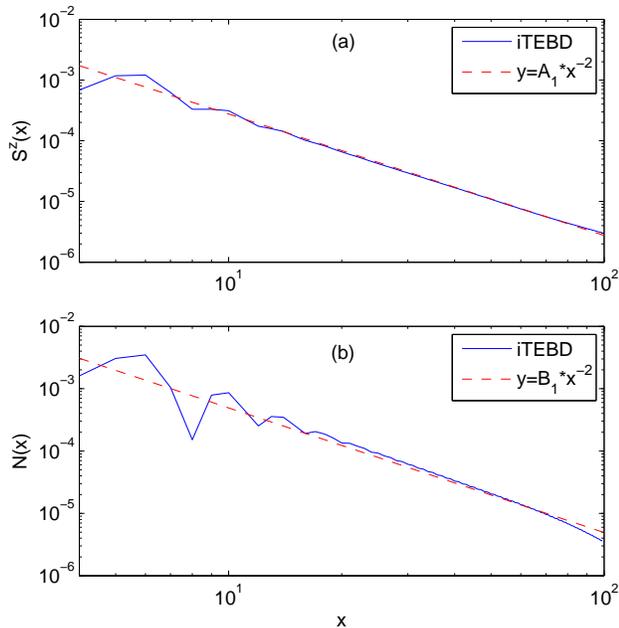}
\caption{(Color online) Real space spin density waves (a) $S^z(x)$ and 
charge density waves (b) $N(x)$ correlations in log-log scale at $U=-2$, $h=1$, 
and $n=1.5$. The red solid lines in (a) and (b) show the asymptotic behavior determined by the bosonization method.} \label{fig:CDW3}
\end{figure}

\begin{figure}[t!]
\includegraphics[width=9cm]{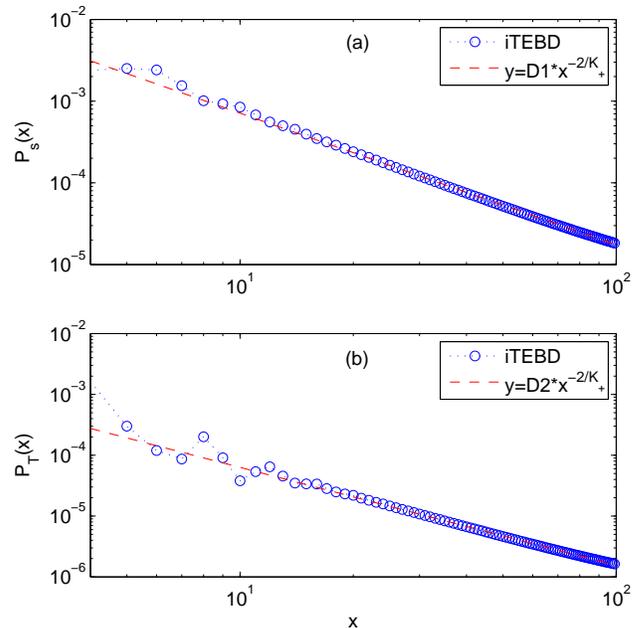}
\caption{(Color online) Real space singlet pairing (a) $P_s(x)$ and triplet pairing (b) $P_T(x)$ correlations in log-log scale at the same parameters as those in Fig.~\ref{fig:CDW3}.
The red dashed lines in (a) and (b) are given by the bosonization method.} \label{fig:SSTS3}
\end{figure}

Not surprisingly, the most divergent susceptibility is that of the pairing sectors.
The singlet and triplet pairing operators have a zero momentum term $e^{i\sqrt{4
\pi}\theta_+}$, which leads to a decay exponent $-2/K_+$ in the real space correlations.
With $K_+>1$ for the attractive interaction we have dominant singlet and triplet 
pairing orders in this phase.

We show the real space correlations of $S^z(x)$ and $N(x)$ in Fig.~\ref{fig:CDW3}(a) and
(b) at $U=-2,\:h=1$ and $n=1.5$ in the log-log scale. The power law functions with exponents 
-2 are also shown for comparison with constants $A_1$ and $B_1$ determined by fitting the data at $x>10$. 
It can be clearly seen that these correlations are well-described by the power law function in Eq.~\ref{eq:Szplus} and Eq.~\ref{eq:Nxplus}. 
The singlet and triplet correlations in real space are aslo demonstrated in Fig.~\ref{fig:SSTS3}.
Both correlation are described by power law functions with the same exponents $-2/K_+$. 
The $K_+=1.24$ is obtained by fitting the $k\to0$ part of the charge density wave structure factor.

The interaction term in the Hamiltonian Eq.~\ref{eq:mainH} under the mean field treatment with the singlet pairing decoupling looks exactly the same as the pairing term induced by the proximity effect.
It can be shown that this region is the topological SC phase in the model with the proximity effect.
To check the topological nature of this phase in the current model, we test the existence of Majorana fermions following the criteria used in Ref.\bibpunct{}{}{;}{n}{,}{,}
\cite{Stoudenmire2011}\bibpunct{}{}{,}{s}{,}{,}. First, Majorana fermions are energy zero modes of the system.
Therefore, the energy difference in even and odd particle number sectors will vanish
if there are Majorana fermions. We scan the ground state energy as a function of
particle numbers $n$ with fixed $h$ and $\mu$ and do not observe any degeneracy in
$n$. 
Second, the entanglement spectrum is found to have a two-fold degeneracy in the topological phase.\cite{Turner2011,Stoudenmire2011} 
We expect it would also be the case if there were Majorana fermions in our system. However, we do not find any evidence of such degeneracy.
Our results suggests that Majorana fermions might not exist in the current system.

\section{Conclusion}
\label{sect:conclusion}

To sum up, we have numerically studied the phase diagram of a spin-orbit coupled interacting Fermi gas in the 1D optical lattice with a transverse magnetic field. 
We find that the intrinsic attractive interaction in the system can not induce the Majorana fermions, unlike the more well-known example in which a pairing is induced by the proximity effect.
Instead, besides a half-filled insulating phase in the strong field limit and a band insulator we identify a LE phase in the weak field and another superconducting phase in the strong field.

The LE phase has dominant singlet and triplet pairing orders. Although they decay with the same exponent, the singlet pairing term has a larger coefficient.
The sub-dominant correlation is that of the charge density wave.
The Luttinger parameters extracted from our numerical results matches reasonably well to those determined by the bosonization method. 
The correlation functions obtained from numerical methods are in good agreement with the asymptotic forms predicted in the bosonization method at strong interaction.
We also find pairing terms with finite momentum.
However, in contrast to the FFLO phase found in the case without the spin-orbit coupling these terms always decay faster than the zero-momentum pairing term.
We compute spin gaps at $h=0$ limit by the DMRG method.
Our results show relatively small spin gaps at weak interaction, which suggest a large correlation length in this region and may cause our failure to capture the correct exponents of the correlation functions in this limit.

The mean field phase at high field region was suggested to be a topological superconducting phase hosting Majorana fermion. 
In our numerical study we again find a superconducting phase with equally dominant singlet and triplet pairing order. 
This phase can be well-described by one Luttinger parameter in all the attractive interaction strength we studied, which agrees with an earlier bosonization study\cite{Fidkowski2011}. 
To search for evidence of Majorana fermions we follow the criteria used in an earlier DMRG study.\cite{Stoudenmire2011} We compute the ground state energy difference in the even and odd parity sectors and the entanglement spectrum at the cut of half the system. However, no supportive evidence is found in our study. It would be an interesting future work to find numerical evidence of the Majorana fermion in a ladder system as suggested by Ref. \bibpunct{}{}{;}{n}{,}{,}\cite{Fidkowski2011} or systems with long range interaction\bibpunct{}{}{;}{s}{,}{,}\cite{Oritz2014}.

\section*{ACKNOWLEDGEMENTS}
We are grateful for the discussion with Ching-Kai Chiu, Professor Kai Sun, Professor Luming Duan, and Professor Jesko Sirker. We especially appreciate Kuei Sun's suggestion on the manuscript.
We acknowledge support by a Thematic Project at Academia Sinica.
Part of the work has been developed by using the DMRG code released within the "Powder with Power" project (www.dmrg.it).

\bibliography{SpinOrbitalBfield,Numerics}

\end{document}